\documentclass[12pt]{iopart}
\usepackage{iopams}  

\usepackage{bm}
\usepackage{mathrsfs}
\usepackage[utf8]{inputenc}
\usepackage{enumerate}
\usepackage{appendix}
\usepackage{dcolumn}
\usepackage{multirow}
\usepackage{float}
\usepackage{color}

\begin{document}
\title[Bespoke analogue space-times: Meta-material mimics]{
\centerline{Bespoke analogue space-times: Meta-material mimics}}
\author{Sebastian Schuster and Matt Visser}
\address{School of Mathematics and Statistics, Victoria University of Wellington, \\ PO Box 600, Wellington 6140, New Zealand}
\ead{sebastian.schuster@sms.vuw.ac.nz, matt.visser@sms.vuw.ac.nz}
\vspace{10pt}
\begin{indented}
\item[]17 January 2018; \LaTeX-ed \today
\end{indented}

\begin{abstract}
Modern meta-materials allow one to construct electromagnetic media with almost arbitrary bespoke permittivity, permeability, and magneto-electric tensors. If (and only if) the permittivity, permeability, and magneto-electric tensors satisfy certain stringent compatibility conditions, can the meta-material be fully described (at the wave optics level) in terms of an effective Lorentzian metric --- an analogue spacetime. We shall consider some of the standard black-hole spacetimes of primary interest in general relativity, in various coordinate systems, and determine the equivalent meta-material susceptibility tensors in a laboratory setting. In static black hole spacetimes (Schwarzschild and the like) certain eigenvalues of the susceptibility tensors will be seen to diverge on the horizon. In stationary black hole spacetimes (Kerr and the like) certain eigenvalues of the susceptibility tensors will be seen to diverge on the ergo-surface.

constitutive tensor, susceptibility tensor, effective metric, analogue spacetime, \\
compatibility conditions.

\end{abstract}


\vspace{2pc}

\maketitle

\def\tr{{\mathrm{tr}}}
\def\cof{{\mathrm{cof}}}
\def\pdet{{\mathrm{pdet}}}

%

\definecolor{purple}{rgb}{1,0,1}
\newcommand{\red}[1]{{\slshape\color{red} #1}}
\newcommand{\blue}[1]{{\slshape\color{blue} #1}}
\newcommand{\purple}[1]{{\slshape\color{purple} #1}}
\hrule
\tableofcontents
\markboth{Bespoke analogue space-times: Meta-material mimics}
{Bespoke analogue space-times: Meta-material mimics}
\bigskip
\hrule
\bigskip
\section{Introduction}
\parindent0pt
\parskip7pt
\vspace{-10pt}

In a recent article~\cite{Schuster:2017} we have re-analyzed and re-explored the question of just when an electromagnetic medium (characterized by permittivity $\epsilon$, permeability $\mu$, and magneto-electric $\zeta$ tensors) is fully equivalent \emph{at the wave optics level} to an effective Lorentzian metric --- an analogue spacetime.  In that article~\cite{Schuster:2017} we explicitly constructed the effective metric in terms of the optical properties. There is a stringent compatibility condition (linking the permittivity, permeability, and magneto-electric tensors) that must be satisfied in order for the analogy to be perfect. (Working within the ray optics approximation is \emph{much} easier.) Ideas along these lines (sometimes only partially implemented) date back, at the very least, to Gordon~\cite{Gordon}, and to Landau and Lifshitz~\cite{Landau-Lifshitz}. There have also been significant related efforts from both the general relativity~\cite{Plebanski:1960,Plebanski:1970,deFelice:1971,Skrotski,Balzas,Anderson} and the optics communities~\cite{Pham,Thompson:2017,Leonhardt:1999}.
These electromagnetic analogue spacetimes complement the acoustic analogue spacetimes  of~\cite{Unruh:1980,Visser:1993,Visser:1997,Barcelo:2000,Barcelo:2003,Visser:2010}. For more general background and history see~\cite{LRR,Visser:2001,Visser:2013}.

\clearpage
In our recent article~\cite{Schuster:2017}, we worked on an arbitrary curved space-time background, and developed a fully relativistic formalism in terms of the 4-velocity of the medium and its $4\times4$ transverse permittivity, permeability, and magneto-electric tensors. For current purposes that would be overkill --- in the current article we are interested in asking what experimental configurations one might need to mimic certain general relativistic black-hole spacetimes (Schwarzschild, Kerr, etcetera) in a laboratory setting. Since all laboratories immediately accessible to us are (for all practical purposes) living in flat Minkowski space, this greatly simplifies the analysis. Since in all laboratories immediately accessible to us there is an obvious ``rest frame'', it is convenient to explicitly split physical laboratory space-time into (space)+(time); this again greatly simplifies the analysis. Since the laboratory spatial slices are flat, it is convenient to adopt Cartesian coordinates in the laboratory, and ``quasi-Cartesian'' coordinates in the spacetime one is trying to mimic.
(If desired a fully relativistic curved background analysis can easily be performed, but that would be superfluous for current purposes.)

\section{General setup}
We shall denote the metric we are trying to mimic by $g_{ab}$, and its inverse by $[g^{-1}]^{ab}$, while the laboratory background metric (Minkowski spacetime) is denoted by $(g_0)_{ab}$ with inverse $(g_0)^{ab}$. Indices will always be raised and lowered using the laboratory metric (which is why we need to use the notation $[g^{-1}]^{ab}$ for the inverse of the metric we want to mimic). We now consider the constitutive tensor~\cite{Schuster:2017} 
\begin{equation}
Z^{abcd} = {1\over2}\;\sqrt{\det(g)\over \det(g_0)} \; 
\left([g^{-1}]^{ac}  [g^{-1}]^{bd} - [g^{-1}]^{ad} [g^{-1}]^{bc} \right),
\label{eq:Z-general}
\end{equation}
which mimics the electromagnetic properties of the metric $g_{ab}$. In (3+1) dimensions this is conformally invariant, so one can at best mimic the metric $g_{ab}$ up to an undetermined conformal factor. (This is an unavoidable aspect of electromagnetism in (3+1) dimensions, and persists at the level of wave optics --- this is a deeper result than the undetermined conformal factor occurring in ray optics and ray acoustics~\cite{LRR}.)  It is convenient to choose this conformal factor so that $\det(g)=\det(g_0)$, 
and so write
\begin{equation}
Z^{abcd} = {1\over2} \; 
\left([g^{-1}]^{ac}  [g^{-1}]^{bd} - [g^{-1}]^{ad} [g^{-1}]^{bc} \right).
\label{eq:Z-specific}
\end{equation}
Normalizing  things this way minimizes the number of tensor densities one has to deal with, and allows one to work with true tensor equations. (If we use Cartesian coordinates for the background Minkowski metric we are effectively setting $\det(g)\to -1$ and can raise and lower spatial indices at will using the Kronecker-delta background spatial metric $[g_0]_{ij} = \delta_{ij}$.) 

The laboratory permittivity, permeability, and magneto-electric tensors are then~\cite{Schuster:2017}:
\begin{equation}
\epsilon^{ij} =  -2 \, Z^{i0j0}; 
\qquad 
[\mu^{-1}]^{ij} =  {1\over2}\;\epsilon^i{}_{kl}\,\epsilon^j{}_{mn}\,Z^{klmn};
\qquad 
\zeta^{ij} = \epsilon^i{}_{kl}\,Z^{klj0}.
\end{equation}
To make this more explicit we first write the metric to be mimicked in (conformally rescaled) Kaluza--Klein inspired form~\cite{Schuster:2017} (this is also sometimes referred to as ``threaded'' form~\cite{Boersma:1994,Bejancu:2015,Gharechahi:2018}):
\begin{equation}
[g^{-1}]^{ab} =  \left[\begin{array}{cc}
-\det(\gamma^{-1}_{\circ\circ}) + \gamma^{-1}_{kl} \beta^k \beta^l & \beta^j\\ 
\beta^i & \gamma^{ij} 
\end{array}\right];  \qquad \det(g)=-1.
\label{eq:g-KKK}
 \end{equation}
Then some relatively easy algebra leads to~\cite{Schuster:2017}:
\begin{equation}
\epsilon^{ij} =  
\left(\gamma^{ij} \{\det(\gamma^{-1}_{\circ\circ}) - \gamma^{-1}_{kl} \beta^k \beta^l \}  
+ \beta^i \beta^j \right);
\qquad
\mu^{ij} =  {\gamma^{ij}\over \det(\gamma^{\circ\circ})};
\label{eq:epsilon-mu-3+1}
\end{equation}
and
\begin{equation}
\zeta^{ij} = -{1\over2}\; \left(  \epsilon^{i}{}_{kl}\beta^l  \gamma^{kj}  \right).
\label{eq:zeta-3+1}
\end{equation}
So we see that this procedure immediately yields the permittivity, permeability, and magneto-electric tensors directly in terms of the metric components ($\gamma^{ij}$ and $\beta^i$) of the (inverse) metric $[g^{-1}]^{ab}$ which we desire to be mimicked. We could proceed with our calculations directly from this step, without further theoretical analysis, but find it useful to first perform some internal consistency checks by looking at the compatibility conditions.

\section{Compatibility conditions}
\subsection{Zero shift}
First suppose that $\beta^i\to0$. This corresponds to the shift vector being zero in the metric to be mimicked. Then one easily sees
\begin{equation}\label{eq:beta=0}
\epsilon^{ij} =  \mu^{ij} =  {\gamma^{ij}\over \det(\gamma^{\circ\circ})};
\qquad\qquad
\zeta^{ij} = 0.
\end{equation}
This very much simplified form of the compatibility conditions is the one that is most often encountered in the literature. From a general relativity perspective this works well if the relativistic spacetime to be mimicked is presented in horizon-non-penetrating coordinates, (such as Schwarzschild curvature coordinates or isotropic coordinates), but is incapable of dealing with horizon penetrating coordinates (such as Painleve--Gullstrand coordinates or Kerr--Schild coordinates or their variants). 

\subsection{Non-zero shift}
Now suppose that $\beta^i\neq0$. It is now convenient to first eliminate $\gamma^{ij}$ from (\ref{eq:epsilon-mu-3+1}) and (\ref{eq:zeta-3+1}), and so deduce
\begin{equation}
\epsilon^{ij} =  	\mu^{ij}\, (1 - \mu^{-1}_{\,kl} \beta^k \beta^l )  + \beta^i \beta^j;
\qquad
\beta^m =   \sqrt{\det(\mu^{\circ\circ})} \; \epsilon^{mk}{}_i \; \mu^{-1}_{\,jk}\; \zeta^{ij}.
\label{eq:beta-neq-0}
\end{equation}

Multiplying the first equation by $[\mu^{-1}]_{jl} \beta^l$ implies $\epsilon^{ij} [\mu^{-1}]_{jl} \beta^l = \beta^i$,  whence
\begin{equation}
[\mu^{-1}]_{jl} \beta^l = [\epsilon^{-1}]_{jl} \beta^l,
\qquad \hbox{and so} \qquad \mu^{-1}_{\,kl} \beta^k \beta^l = \epsilon^{-1}_{\,kl} \beta^k \beta^l.
\end{equation}
This allows us to write
\begin{equation}
 \mu ^{ij} =  {\epsilon^{ij}-  \beta^i \beta^j\over 1 - \epsilon^{-1}_{\,kl} \beta^k \beta^l };
 \qquad
 \det(\mu ^{ij}) = {\det(\epsilon^{ij})\over (1 - \epsilon^{-1}_{\,kl} \beta^k \beta^l)^2}.
 \end{equation}
Here we have used the standard linear algebra result that for any row vector $u$ and any invertible matrix $X$ one has $\det(X+u^T u) = \det(X) \; (1+ u X^{-1} u^T)$. 
Thence
\begin{equation}
{\mu ^{ij}\over\sqrt{\det(\mu^{ij})}} =  {\epsilon^{ij}-  \beta^i \beta^j\over\sqrt{\det(\epsilon^{ij}) }}.
 \end{equation}
 Finally, inverting the relation for $\zeta$ one has
\begin{equation}
\zeta^{ij} 
= -{1\over2}  \sqrt{\det[\mu^{-1}]}\,\epsilon^i{}_{kl} \,\beta^l \mu^{kj} 
= -{1\over2}  \sqrt{\det[\epsilon^{-1}]}\,\epsilon^i{}_{kl} \,\beta^l \epsilon^{kj},
\end{equation}
which also implies
\begin{equation}
\beta^m 
= \epsilon^{mk}{}_{i} \left\{\sqrt{\det(\mu)} [\mu^{-1}]_{kn}  \right\} \zeta^{ni}
= \epsilon^{mk}{}_{i} \left\{\sqrt{\det(\epsilon)} [\epsilon^{-1}]_{kn}  \right\} \zeta^{ni}.
\end{equation}
These various computability and consistency conditions allow for a number of useful internal consistency checks on what could otherwise quite quickly and easily become an impenetrable  forest of  $3\times3$ susceptibility tensors. 

\section{Covariant metric formulation}
Instead of working with the contravariant (inverse) metric $[g^{-1}]^{ab}$ it can be advantageous to work directly with the covariant metric $g_{ab}$.  Matrix inversion quickly leads to either of the two equivalent forms~\cite{Schuster:2017}
\begin{equation}
g_{ab} = 
\left[\begin{array}{cc}
-\sqrt{\det(\mu^{\circ\circ})}^{-1} &  \mu^{-1}_{\,jk}\beta^k\\ 
\mu^{-1}_{\,ik}\beta^k 
&\sqrt{\det(\mu^{\circ\circ})}\left(\mu^{-1}_{\,ij} - ( \mu^{-1}_{\,ik} \beta^k) ( \mu^{-1}_{\,jl}\beta^l)\right)
\end{array}\right],
\label{eq:cov-mu}
\end{equation}	
or alternatively
\begin{equation}
g_{ab} = 
\left[\begin{array}{cc}
-\sqrt{\det(\epsilon^{\circ\circ})}^{-1}  (1 - \epsilon^{-1}_{\,kl} \beta^k \beta^l) &  \epsilon^{-1}_{\,jk}\beta^k\\ 
\epsilon^{-1}_{\,ik}\beta^k 
&\sqrt{\det(\epsilon^{\circ\circ})} \left( \epsilon^{-1}_{ij}  \right)
\end{array}\right].
\label{eq:cov-epsilon}
\end{equation}
Here again
\begin{equation}
\beta^m 
= \epsilon^{mk}{}_{i} \left\{\sqrt{\det(\mu)} [\mu^{-1}]_{kn}  \right\} \zeta^{ni}
= \epsilon^{mk}{}_{i} \left\{\sqrt{\det(\epsilon)} [\epsilon^{-1}]_{kn}  \right\} \zeta^{ni},
\end{equation}
and 
\begin{equation}
\zeta^{ij} 
= -{1\over2}  \sqrt{\det[\mu^{-1}]}\epsilon^i{}_{kl} \beta^l \mu^{kj} 
= -{1\over2}  \sqrt{\det[\epsilon^{-1}]}\epsilon^i{}_{kl} \beta^l \epsilon^{kj},
\end{equation}
subject to 
\begin{equation}
\epsilon^{ij} =  	\mu^{ij}\, (1 - \mu^{-1}_{\,kl} \beta^k \beta^l )  + \beta^i \beta^j;
\qquad
 \mu ^{ij} =  {\epsilon^{ij}-  \beta^i \beta^j\over 1 - \epsilon^{-1}_{\,kl} \beta^k \beta^l }.
\label{eq:compatability}
\end{equation}
Any one of these various routes can be used to map the spacetime metric to be mimicked into a triad of equivalent laboratory permittivity, permeability, and magneto-electric tensors.
We shall now consider some specific examples.

\section{Schwarzschild spacetime}
The Schwarzschild spacetime is one of the first (and one of the most important) of the known exact solutions in Einstein's general relativity. As such it is an excellent test-bed for any analogue space-time programme.

\subsection{Cartesian curvature coordinate form}
We wish to fit the curvature coordinate form (sometimes called the Hilbert form) of the Schwarzschild spacetime metric
\begin{equation}
ds^2 = - (1-2m/r)dt^2 + {dr^2\over1-2m/r} + r^2 d\Omega^2
\label{eq:hilbert}
\end{equation}
to the formalism developed above.
It is convenient to first adopt a ``Cartesian'' version of curvature coordinates by defining
\begin{equation}
x^a = (t,x,y,z) = (t,r\sin\theta\cos\phi,r\sin\theta\sin\phi,r\cos\theta).
\end{equation}
Now write
\begin{equation}
\hat r_i = \left({x\over r}, {y\over r}, {z\over r} \right),
\end{equation}
and define the projection operator $P_{ij} = \delta_{ij} - \hat r_i \hat r_j$.  Then 
 \begin{equation}
ds^2 = g_{ab} dx^a dx^b=  - (1-2m/r)dt^2 + {(\hat r_i \; dx^i)^2\over1-2m/r} + P_{ij} dx^i dx^j;
\label{eq:pseudo-cartesian}
\end{equation}
with $\det(g_{ab}) = -1$. This casts the Schwarzschild spacetime in curvature coordinates into a quasi-Cartesian form suitable for direct comparison with laboratory quantities. 

(If desired, one could equally stay in the $(t,r,\theta,\phi)$ coordinates of equation (\ref{eq:hilbert}), but then one would want to write the laboratory metric $[g_0]_{ab}$ in spherical polar coordinates so as to have $\det(g_{ab})=\det([g_0]_{ab})$. In computations, one would then need to keep track of various components of the background metric $[g_0]_{ab}$ to raise and lower indices. This is unnecessarily indirect for current purposes.)

Starting, for instance, from equation (\ref{eq:cov-mu}) applied to equation (\ref{eq:pseudo-cartesian}) there are three obvious deductions:
\begin{equation}
\sqrt{\det(\mu^{\circ\circ})}^{-1} = 1-2m/r; \qquad\qquad \beta^i  = 0;
\end{equation}
and
\begin{equation}
 \sqrt{\det(\mu^{\circ\circ})} \; \mu^{-1}_{\,ij} =  {\hat r_i \; \hat r_j\over1-2m/r} + P_{ij}.
 \end{equation}

 This is easily solved to yield
 \begin{equation}
 \mu^{-1}_{\,ij}  = (1-2m/r)P_{ij} + \hat r_i \hat r_j;
 \qquad
 \mu_{\,ij}  = (1-2m/r)^{-1} P_{ij} + \hat r_i \hat r_j.
 \end{equation} 
This satisfies the determinant condition above, and in summary we have 
 \begin{equation}
 \epsilon_{\,ij} = \mu_{\,ij}  = (1-2m/r)^{-1} P_{ij} + \hat r_i \hat r_j; \qquad\qquad \zeta_{ij} = 0.
 \end{equation} 
These are the equivalent electric and magnetic properties to the Schwarzschild geometry in curvature-coordinate form.
We can if desired rewrite this as 
\begin{equation}
\fl
\epsilon_{rr} = \mu_{rr} = 1; \qquad\quad
\epsilon_{\hat\theta\hat\theta}=\epsilon_{\hat\phi\hat\phi} = \mu_{\hat\theta\hat\theta}=\mu_{\hat\phi\hat\phi} 
= (1-2m/r)^{-1}  > 1; \quad\quad \zeta=0.
\end{equation}
Note $\epsilon=\mu \geq 1$, which is ``physically reasonable", and that both $\epsilon$ and $\mu$ diverge at the horizon. 

Having both $\epsilon\geq 1$ and $\mu \geq 1$, or more precisely the eigenvalues of $\epsilon_{ij}$ and $\mu_{ij}$ greater than unity, is the easy ``standard case'' for permittivity and permeability.  There are certainly 
more \emph{outr\'e} situations where $\epsilon\leq 1$ and $\mu \leq 1$, or even $\epsilon\leq 0$ and $\mu \leq 0$, but we shall see that we do not need to appeal to exotic media of that type to simulate Schwarzschild black holes. (Nor for that matter will we need to, or even want to,  resort to complex permittivity and permeability.) 

The fact that some optical properties must diverge at the horizon can be back-tracked, at least, to work by Reznik~\cite{Reznik:1997}. Reznik considered the specific case where (in the laboratory frame) $\epsilon=\mu=1/(\alpha z)$ and $\zeta=0$, and showed that this is equivalent to an effective Rindler spacetime $ds^2 =  -\alpha^2 z^2dt^2 + dx^2 + dy^2 + dz^2$ with Rindler horizon at $z=0$. The point is that trapping electro-magnetic radiation, in a frequency independent manner, will require something odd to happen at the horizon.
(For stationary black holes [Kerr and the like] the divergences in the susceptibility tensors will be seen to shift to the ergo-surface.)

\subsection{Cartesian Painleve--Gullstrand form}\label{sec:sch-PG}
We now need to fit the spacetime metric 
\begin{equation}
\fl
\qquad
ds^2 = g_{ab} dx^a dx^b = - dt^2 + \delta_{ij} \left(dx^i - \sqrt{2m/r} \; \hat r_i \,dt\right)  \left(dx^j - \sqrt{2m/r} \; \hat r_j \,dt\right)
\label{eq:pg}
\end{equation}
to the formalism developed above. (Observe that this is already in quasi-Cartesian form and that $\det(g_{ab})=-1$.)
Starting, for instance, from equation (\ref{eq:cov-mu}), there are three obvious deductions:
\begin{equation}
\sqrt{\det(\mu^{\circ\circ})}^{-1} = 1-2m/r; \qquad \mu^{-1}_{ik}  \beta^k  = \sqrt{2m/r} \; \hat r_i;
\end{equation}
and
\begin{equation}
\delta_{ij} = \sqrt{\det(\mu^{\circ\circ})}\left(\mu^{-1}_{\,ij} 
- \left[\sqrt{2m/r} \; \hat r_i\right]\left[\sqrt{2m/r} \; \hat r_j\right]\right).
 \end{equation}
 Thus our demand for the fulfilment of the consistency conditions leads to
 \begin{equation}
 \mu^{-1}_{\,ij}  = (1-2m/r)\delta_{ij} + (2m/r) \hat r_i \hat r_j.
\end{equation} 
Thence, using the same projection operator $P_{ij} = \delta_{ij} - \hat r_i \hat r_j$ as in the previous section,  we see
\begin{equation}
 \mu^{-1}_{\,ij}  = (1-2m/r)P_{ij} +  \hat r_i \hat r_j;
 \qquad
\mu_{\,ij}  = (1-2m/r)^{-1} P_{ij} +  \hat r_i \hat r_j.
 \end{equation} 
This satisfies the determinant condition above and we now have 
 \begin{equation}
 \mu_{\,ij}  = (1-2m/r)^{-1} P_{ij} + \hat r_i \hat r_j; \qquad \zeta_{ij} = -{1\over2} \sqrt{2m\over r} \varepsilon_{ijk} \hat r^k .
 \end{equation} 
Note that $ \mu_{\,ij} $ is the same as for curvature coordinates, though $\zeta_{ij}$ differs.
There is a good reason for this: the curvature and Painleve--Gullstrand coordinates are related by simple coordinate transformation of the form $t\to t + f(r)$.
The $\epsilon$-tensor can be calculated in several ways, for instance
\begin{equation}
\epsilon^{ij} =  
	\mu^{ij}\, (1 - \mu^{-1}_{\,kl} \beta^k \beta^l )  + \beta^i \beta^j = 
	\mu^{ij} (1-2m/r) + (2m/r) \hat r^i \hat r^j = \delta^{ij}.
\end{equation}	
This is simply the identity matrix. This could also be extracted from equation (\ref{eq:cov-epsilon}) where matching to the Painleve--Gullstrand form (\ref{eq:pg}) of the metric 
leads to the same results.

\bigskip
\noindent
In summary, for the Schwarzschild geometry in Painleve--Gullstrand coordinates we have
\begin{equation}
\epsilon_{\,ij} = \delta_{ij};
\quad
 \mu_{\,ij}  = (1-2m/r)^{-1} P_{ij} + \hat r_i \hat r_j; 
 \quad 
 \zeta_{ij} = -{1\over2} \sqrt{2m\over r} \varepsilon_{ijk} \hat r^k .
 \end{equation} 
Note $\epsilon=1$ and $\mu \geq 1$, which is ``physically reasonable", and that $\mu$ diverges at the horizon.
 
\subsection{Cartesian Kerr--Schild form}\label{sec:sch-KS}

We now need to fit
\begin{equation}
g_{ab} = \eta_{ab} + {2m\over r} \ell_a \ell_b; \qquad   \ell_a = ( -1; \hat r_i);
\qquad \det(g_{ab})=-1.
\end{equation}
From equation (\ref{eq:cov-mu}), three obvious deductions are these:
\begin{equation}
\sqrt{\det(\mu^{\circ\circ})}^{-1} = 1-2m/r; \qquad \mu^{-1}_{ik}  \beta^k  = (2m/r) \; \hat r_i;
\end{equation}
and
\begin{equation}
\delta_{ij} + (2m/r)\; \hat r_i\; \hat r_j = 
\sqrt{\det(\mu^{\circ\circ})}\left(\mu^{-1}_{\,ij} - \left[(2m/r) \; \hat r_i\right]\left[(2m/r) \; \hat r_j\right]\right).
 \end{equation}
 Then we want
 \begin{equation}
 \mu^{-1}_{\,ij} = (1-2m/r)[\delta_{ij} +  (2m/r)\; \hat r_i\; \hat r_j ] + [(2m/r) \; \hat r_i][(2m/r) \; \hat r_j].
 \end{equation}
 That is
 \begin{equation}
 \mu^{-1}_{\,ij}  = (1-2m/r)\delta_{ij} + (2m/r) \hat r_i \hat r_j.
 \end{equation} 
 Thence
 \begin{equation}
 \mu^{-1}_{\,ij}  = (1-2m/r)P_{ij} +  \hat r_i \hat r_j;
 \qquad
\mu_{\,ij}  = (1-2m/r)^{-1} P_{ij} +  \hat r_i \hat r_j.
 \end{equation} 
This satisfies the determinant condition above, and we have 
 \begin{equation}
 \mu_{\,ij}  = (1-2m/r)^{-1} P_{ij} + \hat r_i \hat r_j; 
 \qquad
 \zeta_{ij} = -{m\over r} \,\varepsilon_{ijk} \,\hat r^k .
 \end{equation} 
Again $\mu_{\,ij} $ is the same as for curvature coordinates, and Painleve--Gullstrand coordinates, only $\zeta_{ij}$ differs.
The same fundamental reason underlies this observation: Ultimately, all three of the curvature, Painleve--Gullstrand, and Kerr--Schild coordinates are related by simple coordinate transformations of the form $t\to t + f(r)$.
For the $\epsilon$-tensor we can easily calculate
\begin{eqnarray}
\epsilon^{ij} &=&  
	\mu^{ij}\, (1 - \mu^{-1}_{\,kl} \beta^k \beta^l )  + \beta^i \beta^j =
	\mu^{ij} (1-[2m/r]^2) + (2m/r)^2 \hat r^i \hat r^j \nonumber\\
	&=& (1+2m/r) P^{ij} +\hat r^i \hat r^j.
\end{eqnarray}	
Alternatively, starting from equation (\ref{eq:cov-epsilon}), and fitting the Kerr--Schild form of the Schwarzschild metric, we can obtain the same results.

\bigskip
\noindent
In summary, for the Schwarzschild metric in Cartesian Kerr--Schild form we have
\begin{equation}
\fl 
\epsilon_{\,ij} = (1+2m/r)  P_{ij}+ \hat r_i\; \hat r_j;
\quad
\mu_{\,ij}  = (1-2m/r)^{-1} P_{ij} + \hat r_i \hat r_j; 
\quad 
\zeta_{ij} = -{m\over r} \varepsilon_{ijk} \hat r^k .
\end{equation}
Note $\epsilon\geq1$ and $\mu \geq 1$, which is ``physically reasonable", and that $\mu$ (but not $\epsilon$) diverges at the horizon. 

\subsection{Cartesian Gordon form}\label{sec:sch-gordon}

The ``Gordon form'' of the Schwarzschild metric~\cite{Rosquist,Liberati} is less well-known than perhaps is should be.
We now consider
\begin{equation}
g_{ab} =\sqrt{n} \left( \eta_{ab} + [1-n^{-2}]V_a V_b\right); \quad   
V_a = \left(- \sqrt{1+2m/r}; \sqrt{2m/r} \; \hat r_i\right).
\end{equation}
Here $n$ is an arbitrary constant, $V_a$ is a 4-velocity, and the parameter $m$ is proportional to the physical mass of the Schwarzschild spacetime.
The overall conformal factor $\sqrt{n}$ in the metric enforces $\det(g_{ab})=-1$. 
It is easy to check that this metric is Ricci flat. Here $n$ can be interpreted as the refractive index (in its rest frame) of some medium with 4-velocity $V_a$. To interpret the parameter $m$ note
\begin{equation}
\fl\qquad
g_{tt} = - \sqrt{n}(1-[1-n^{-2}] (1+2m/r)) =  - \sqrt{n} \left(n^{-2} - [1-n^{-2}] (2m/r)\right).
\end{equation}
That is
\begin{equation}
\fl\qquad
g_{tt} =  - n^{-3/2} \left(1 - [n^{2}-1] (2m/r)\right) \propto - \left(1 - [n^{2}-1] (2m/r)\right) ,
\end{equation}
whence one can read off the physical mass as
\begin{equation}
{G_\mathrm{Newton} \, m_\mathrm{physical}\over c^2} = (n^2-1) m.
\end{equation}
For further discussion see references~\cite{Rosquist,Liberati}.

Now in the rest frame $V$ of the medium, $\epsilon_V = \mu_V =  n$ are (by assumption) isotropic, and $\zeta_V=0$; but we are chiefly interested in calculating the permittivity, permeability, and magneto-electric tensors \emph{in the laboratory fame}.\,\footnote{To proceed, we could in principle perform a ``local'' Lorentz transformation from the rest frame $V$ of the medium into the laboratory frame --- along the lines of Appendix B of reference~\cite{Schuster:2017},
but the ``matching'' analysis presented here is more straightforward.}

Matching to equation (\ref{eq:cov-mu}) we have three obvious deductions:
\begin{equation}
\sqrt{\det(\mu^{\circ\circ})}^{-1} = \sqrt{n}\left\{1-[1-n^{-2}] (1+2m/r)\right\}; 
\end{equation}
\begin{equation}
 \mu^{-1}_{ik}  \beta^k  = \sqrt{n} [1-n^{-2}] \sqrt{(1+2m/r)(2m/r)}\; \hat r_i;
\end{equation}
and 
\begin{eqnarray}
\sqrt{\det(\mu^{\circ\circ})}&\left(\mu^{-1}_{\,ij} - \left[n(1-n^{-2})^2(1+2m/r)(2m/r)\hat r_i \hat r_j\right]\right)
\nonumber\\[5pt]
&=
\sqrt{n}\left\{ \delta_{ij} + [1-n^{-2}](2m/r) \hat r_i \hat r_j \right\}.
 \end{eqnarray}
 Multiply the first of these equations by the third:
 \begin{eqnarray}
 &&
 \left(\mu^{-1}_{\,ij} - \left[n(1-n^{-2})^2(1+2m/r)(2m/r)\hat r_i \hat r_j\right]\right)
 \nonumber
 \\[5pt]
&&\qquad =
n \left\{1-[1-n^{-2}] (1+2m/r)\right\} \left\{ \delta_{ij} + [1-n^{-2}](2m/r) \hat r_i \hat r_j \right\}.
\end{eqnarray}
Therefore
\begin{eqnarray}
\mu^{-1}_{\,ij} 
&=&
n \left\{1-[1-n^{-2}] (1+2m/r)\right\} \left\{ \delta_{ij} + [1-n^{-2}](2m/r) \hat r_i \hat r_j \right\}
\nonumber\\
&&
+
\left[n(1-n^{-2})^2(1+2m/r)(2m/r)\hat r_i \hat r_j\right].
\end{eqnarray}
This simplifies to
\begin{equation}
\mu^{-1}_{\,ij} 
= n \left\{1-[1-n^{-2}] (1+2m/r)\right\}  P_{ij} + n^{-1}\; \hat r_i \hat r_j,
\end{equation}
implying
\begin{equation}
\mu_{\,ij} 
= {P_{ij} \over n \left\{1-[1-n^{-2}] (1+2m/r)\right\} } + n\; \hat r_i \hat r_j.
\end{equation}
Note that at large $r$
\begin{equation}
\mu^{-1}_{\,ij}  \to n^{-1} \delta_{ij}; \qquad \mu_{ij} \to n \; \delta_{ij}. 
\end{equation}

To calculate the $\zeta$-tensor it is useful to first note:
\begin{equation}
\sqrt{\det(\mu^{-1}_{\,ij})}
= \sqrt{n} \left\{1-[1-n^{-2}] (1+2m/r)\right\};
\end{equation}
\begin{equation}
 \mu^{-1}_{ik}  \beta^k  = n^{-1}  \beta_i ;
 \end{equation}
 \begin{equation}
 \beta_i =  n^{3/2} [1-n^{-2}] \sqrt{(1+2m/r)(2m/r)}\; \hat r_i.
 \end{equation}
 Thence
\begin{eqnarray}
\zeta^{ij} 
&=& -{1\over2}  \sqrt{\det[\mu^{-1}]}\epsilon^i{}_{kl} \beta^l \mu^{kj} 
\\
&=& -{1\over2}  \left( [n^{2}-1] \sqrt{(1+2m/r)(2m/r)} \right)\epsilon^{ij}{}_k \; \hat r^k .
\end{eqnarray}
So $\zeta^{ij} \to  0$ at large $r$.

 To calculate the $\epsilon$-tensor it is useful to first note:
 \begin{eqnarray}
 \beta_i \beta_j&=&  n^{3} [1-n^{-2}]^2 (1+2m/r)(2m/r)\; \hat r_i \hat r_j 
 \\
 &=& 
  n^{-1} [n^{2}-1]^2 (1+2m/r)(2m/r)\; \hat r_i \hat r_j; 
 \end{eqnarray}
 whence
 \begin{equation}
 \mu^{-1}_{ik}  \beta_i \beta^k =n^{-2} [n^2-1]^2(1+2m/r)(2m/r).
 \end{equation}
Then we have
\begin{eqnarray}
\epsilon^{ij} &=&  
\mu^{ij}\, \left(1 - \mu^{-1}_{\,kl} \beta^k \beta^l \right)  + \beta^i \beta^j 
\\
&=& 
\mu^{ij} \left(1-n^{-2} [n^2-1]^2(1+2m/r)(2m/r)\right) 
\nonumber\\ &&
+  n^{-1}[n^2-1]^2(1+2m/r)(2m/r) \hat r^i \hat r^j 
\\
&=& {(1-n^{-2}[n^2-1]^2(1+2m/r)(2m/r)) \over n \left\{1-[1-n^{-2}] (1+2m/r)\right\}} P^{ij} 
\\
&&\quad+
\Big( n  (1-n^{-2}[n^2-1]^2(1+2m/r)(2m/r))  
\nonumber\\&&
\qquad + (n^{-1}[n^2-1]^2(1+2m/r)(2m/r)) \Big) \hat r^i \hat r^j.
\end{eqnarray}
Simplifying
\begin{equation}
\epsilon^{ij} =  
	n \{ 1+[1-n^{-2}] (2m/r) \} P^{ij} 
	+ n \hat r^i \hat r^j.
\end{equation}
Note that at large $r$
\begin{equation}
\epsilon^{-1}_{\,ij}  \to n^{-1} \delta_{ij}; \qquad \epsilon_{ij} \to n \; \delta_{ij}. 
\end{equation}
Alternatively, we could also extract the same results from equation (\ref{eq:cov-epsilon}), or by performing a ``local'' Lorentz transformation from the rest frame $V$ of the medium into the laboratory frame.

\bigskip
\noindent
In summary, for the Schwarzschild spacetime in Gordon form
\begin{equation}
\epsilon_{ij} =  n \{ 1+[1-n^{-2}] (2m/r) \} P_{ij}  + n \; \hat r_i \hat r_j;
\end{equation}
\begin{equation}
\mu_{\,ij}  = {P_{ij} \over n \left\{1-[1-n^{-2}] (1+2m/r)\right\} } + n\; \hat r_i \hat r_j;
\end{equation}
\begin{equation}
\zeta_{ij} 
= -{1\over2}  \left( [n^{2}-1] \sqrt{(1+2m/r)(2m/r)} \right)\epsilon_{ijk} \; \hat r^k .
\end{equation}
Note $\epsilon\geq1$ and $\mu \geq 1$, which is ``physically reasonable", and that $\mu$ (but not $\epsilon$) diverges at the horizon, (which is located by solving $1-[1-n^{-2}] (1+2m/r)=0$, that is $r_H=2(n^2-1)m$).

\subsection{Cartesian isotropic form}\label{sec:sch-iso}

We would now need to fit
\begin{equation}
g_{ab} = - \left(1-{m\over2r}\over1+{m\over2r}\right)^2 dt^2 
+ \left(1+{m\over2r}\right)^4 |d\vec x|^2; \qquad  r = |\vec x|. 
\end{equation}
But note that curvature, Painleve--Gullstrand, and Kerr--Schild coordinates, when put in quasi-Cartesian form, all have the nice property $\det(g_{ab})=-1$, whereas the isotropic form of the metric does not share this property.

Since our electromagnetic  effective metrics were, (thanks to conformal invariance of electromagnetism in 3+1 dimensions), all chosen to satisfy $\det(g)=-1$, for isotropic coordinates we should pull out an overall conformal factor and write 
\begin{equation}
\fl
g_{ab} =  \sqrt[4]{  \left(1-{m\over2r}\right)^2\,\left(1+{m\over2r}\right)^{10}} \; 
\left(
-\sqrt[4]{\left(1-{m\over2r}\right)^6\over\left(1+{m\over2r}\right)^{18}} \; dt^2 
+ \sqrt[4]{\left(1+{m\over2r}\right)^6\over\left(1-{m\over2r}\right)^2} \; |d\vec x|^2
\right).
\end{equation}
Discard the overall conformal factor,  since electromagnetism is conformally invariant in 3+1 dimensions, focus on what remains. Then we need to fit (see for instance~\cite{deFelice:1971})
\begin{equation}
\fl
g_{ab} = 
\left(
-\sqrt[4]{\left(1-{m\over2r}\right)^6\over\left(1+{m\over2r}\right)^{18}} \; dt^2 
+ \sqrt[4]{\left(1+{m\over2r}\right)^6\over\left(1-{m\over2r}\right)^2}\;  |d\vec x|^2
\right) 
=
- B^{-6} \; dt^2 + B^2 \; |d\vec x|^2.
\end{equation}
Three obvious deductions are:
\begin{equation}
\sqrt{\det(\mu^{\circ\circ})}^{-1} = B^{-6}; 
\qquad
\beta^k  = 0;
\qquad 
B^2 \delta_{ij} = 
\sqrt{\det(\mu^{\circ\circ})}\;\mu^{-1}_{\,ij}.
 \end{equation}
 Then we want
 \begin{equation}
 \mu^{-1}_{\,ij} = B^{-4} \delta_{ij};
 \qquad
 \mu_{\,ij}  = B^{+4} \delta_{ij} .
 \end{equation} 
This satisfies the determinant condition and we have 
 \begin{equation}
 \epsilon_{\,ij} = \mu_{\,ij}  = B^4 \delta_{ij}; \qquad \zeta_{ij} = 0.
 \end{equation} 
 For the Schwarzschild geometry in isotropic coordinates
 \begin{equation}
 B^4 =  \sqrt{\left(1+{m\over2r}\right)^6\over\left(1-{m\over2r}\right)^2} 
 = {\left(1+{m\over2r}\right)^3\over\left|1-{m\over2r}\right|}> 1;
 \end{equation} 
 Since $\epsilon=\mu>1$ this is ``physically appropriate''. 
 Explicitly
 \begin{equation}
 \epsilon_{\,ij} = \mu_{\,ij}  =  {\left(1+{m\over2r}\right)^3\over\left|1-{m\over2r}\right|}\;\delta_{ij}; \qquad \zeta_{ij} = 0.
\end{equation}
That is, now in terms of a position-dependent refractive index $n(r)$, (as measured in the laboratory), we have: 
\begin{equation}
n(r) =\epsilon(r) = \mu(r) =  {\left(1+{m\over2r}\right)^3\over\left|1-{m\over2r}\right|}; \qquad \zeta = 0.
\end{equation}
Note the divergence in the laboratory optical parameters at the horizon.
Mathematically, the situation for isotropic coordinates is now (because of the need for an explicit conformal factor) qualitatively 
different than the situation for curvature coordinates, Painleve--Gullstrand coordinates, and Kerr--Schild coordinates.
Another reason for not worrying about any overall conformal factor is this: Jacobson and Kang have shown that under suitable regularity conditions the surface gravity (and hence the Hawking temperature) is a conformal invariant~\cite{Jacobson}.

\section{Static spherically symmetric spacetimes}\label{sec:ssss}

Merely by invoking spherical symmetry we have
\begin{equation}
\epsilon_{ij} = \epsilon_\perp P_{ij} + \epsilon_\parallel \; \hat r_i \hat r_j;
\qquad\;
\det(\epsilon_{ij}) =  \epsilon_\perp^2\; \epsilon_\parallel;
\end{equation}
\begin{equation}
\mu_{ij} = \mu_\perp P_{ij} + \mu_\parallel \; \hat r_i \hat r_j;
\qquad
\det(\mu_{ij}) =  \mu_\perp^2\; \mu_\parallel;
\end{equation}
and
\begin{equation}
\beta^i =  \beta \; \hat r^i.
\end{equation}
When added to the compatibility conditions, this severely constrains the susceptibility tensors. Observe that  $\epsilon^{ij} = \mu^{ij}\, (1 - \mu^{-1}_{\,kl} \beta^k \beta^l )  + \beta^i \beta^j$ implies both
\begin{equation}
\epsilon_\perp = \mu_\perp \left(1- {\beta^2\over\mu_\parallel}\right); 
\qquad \hbox{and} \qquad
\epsilon_\parallel = \mu_\parallel.
\end{equation}
Then for the magneto-electric tensor
\begin{equation}
\zeta^{ij} 
= -{1\over2}  \sqrt{\det[\mu^{-1}]}\epsilon^i{}_{kl} \beta^l \mu^{kj} 
= -{1\over2}  {\beta\over\sqrt{\mu_\parallel}} \; \epsilon^{ij}{}_k \; \hat r^k .
\end{equation}
In summary, spherical symmetry by itself is enough to imply
\begin{equation}
\fl
\epsilon_{ij} =  \mu_\perp \left(1- {\beta^2\over\mu_\parallel}\right) P_{ij} + \mu_\parallel \; \hat r_i \hat r_j;
\quad
\mu_{ij} = \mu_\perp P_{ij} + \mu_\parallel \; \hat r_i \hat r_j;
\quad 
\zeta^{ij} = -{1\over2}  {\beta\over\sqrt{\mu_\parallel}} \; \epsilon^{ij}{}_k \; \hat r^k .
\end{equation}
It is easy to check that our previous computations for Painleve--Gullstrand, Kerr--Schild, and Gordon forms of the Schwarzschild metric satisfy these conditions.

\section{Kerr spacetime}\label{sec:kerr}
The Kerr spacetime is particularly interesting and important --- representing as it does the astrophysically relevant case of a rotating black hole. (See for instance~\cite{Kerr-book,Kerr-intro}.) We shall analyze the Cartesian Kerr--Schild and Cartesian Doran forms of the metric --- 
in fact the analysis is more general than just Kerr itself, and will apply to any spacetime that can be cast into  Cartesian Kerr--Schild, or Cartesian Doran form.
\subsection{Cartesian Kerr--Schild form}

Let us first consider the Kerr spacetime in Cartesian Kerr--Schild coordinates. (See for instance~\cite{Kerr-book,Kerr-intro}.) We need to fit:
\begin{equation}
\fl
\qquad
g_{ab}  = \eta_{ab} + 2\Phi \ell_a \ell_b;   \qquad \ell_a = (1;\ell_i);  \qquad ||\ell_i||=1;
\qquad \det(g_{ab})=-1.
\end{equation}
Here $\ell^a$ is the Kerr null congruence, $\ell^i$ is a unit vector in 3-space, and $\Phi$ is the gravitational potential. We will not need any more detailed information. (Therefore the present analysis is applicable, with only trivial modifications,  to any spacetime that can be represented in Kerr--Schild form.)
Starting from equation (\ref{eq:cov-mu}) there are two obvious deductions:
\begin{equation}
\sqrt{\det(\mu^{\circ\circ})}^{-1} = 1-2\Phi; \qquad\qquad \mu^{-1}_{\,jk}\beta^k  = 2\Phi \ell_j .
\end{equation}
Now write the space part of the metric as:
 \begin{equation}
 \left(\delta_{ij} +2\Phi \ell_i \ell_j \right) 
 =
 \sqrt{\det(\mu^{\circ\circ})} (\mu^{-1}_{\,ij} - 4\Phi^2 \ell_i \ell_j).
 \end{equation}
 Thence
 \begin{equation}
 \mu^{-1}_{\,ij}  = (1-2\Phi)(\delta_{ij} +2\Phi \ell_i \ell_j )  +4\Phi^2 \ell_i \ell_j=
 (1-2\Phi)(\delta_{ij}) - 2\Phi \ell_i \ell_j.
 \end{equation} 
 Define a new projection operator $P_{ij} = \delta_{ij} - \ell_i \ell_j$, (now slightly different from that used in the spherically symmetric case because it uses the unit vector $\ell_i$ rather than $\hat r_i$). Then
 \begin{equation}
 \mu^{-1}_{\,ij}  = (1-2\Phi)P_{ij} +\ell_i \ell_j,
 \end{equation} 
 and so
  \begin{equation}
 \mu_{\,ij}  = (1-2\Phi)^{-1} P_{ij} + \ell_i \ell_j = {\delta_{ij} - 2\Phi \ell_i\ell_j\over 1-2\Phi},
 \end{equation} 
 while
 \begin{equation}
 \beta^k =2 \Phi \ell^i.
 \end{equation}
 Furthermore
 \begin{equation}
 \det(\mu^{-1}_{\,ij}  ) = (1-2\Phi)^2.
 \end{equation}
 This satisfies the determinant condition above, and we have 
\begin{eqnarray}
\epsilon^{ij} &=&  
\mu^{ij}\, (1 - \mu^{-1}_{\,kl} \beta^k \beta^l )  + \beta^i \beta^j\\
&=&
( (1-2\Phi)^{-1} P_{ij} + \ell_i \ell_j) (1-4\Phi^2)+ 4\Phi^2 \ell_i \ell_j.
\end{eqnarray}
Therefore
\begin{equation}
\epsilon^{ij} = 
P_{ij} (1+2\Phi) +  \ell_i \ell_j;
\qquad
\det(\epsilon^{ij}) = (1+2\Phi)^2.
\end{equation}
Now
\begin{equation}
 [\epsilon^{-1}]_{ij} = {P_{ij} \over (1+2\Phi)} +  \ell_i \ell_j = {\delta_{ij}+ 2\Phi \ell_i \ell_j\over 1+2\Phi},
 \end{equation}
so we have the interesting observation that in terms of the 3-metric
\begin{equation}
  [\epsilon^{-1}]_{ij} = {g_{ij}\over \det(g_{ij})}.
  \end{equation}
Finally, for the magneto-electric tensor
\begin{eqnarray}
  \zeta^{ij} &=& -{1\over2}\;
\left(  \epsilon^{i}{}_{kl}\beta^l  \mu^{kj} \over \sqrt{\det(\mu^{\circ\circ})} \right)
\\
&=& -{1\over2}\;\left(  [1-2\Phi] \epsilon^{i}{}_{kl} [2\Phi\ell^l] \left[{\delta^{kj} - 2\Phi \ell^k\ell^j\over 1-2\Phi}\right]\right) = - \Phi \epsilon^{ij}{}_{l} \ell^l.
\end{eqnarray}
That is
\begin{equation}
\zeta^{ij} =
 - \Phi \epsilon^{ij}{}_{l} \ell^l.
\end{equation}
We can get exactly the same results by instead starting from equation (\ref{eq:cov-epsilon}).

\bigskip
\noindent
 In summary, for the Kerr spacetime (in Cartesian Kerr--Schild coordinates) we have:
 \begin{equation}
 \epsilon^{ij} =  (1+2\Phi) P^{ij} +  \ell^i\ell^j;
 \quad
\mu^{ij} 
=
{P^{ij} \over 1-2\Phi} + \ell^i\; \ell^j;
\quad
\zeta^{ij} =
 - \Phi \epsilon^{ij}{}_{l} \ell^l.
\end{equation}
Under usual conditions $\Phi\geq 0$ so  the eigenvalues of $\epsilon_{ij}$ and $\mu_{ij}$ are greater than or equal to unity. This is ``physically appropriate''.  Two of the eigenvalues of $\mu_{ij}$ diverge when $2\Phi=1$, which corresponds to $g_{00}=0$, which defines the ergo-surface, not the horizon. Since a laboratory observer goes ``superluminal'' in the effective metric at the ergo-surface, and since one expects to see super-radiance from behind the ergo-surface, it should not be all that surprising that the optical properties diverge there.

\medskip
To be explicit, for Kerr spacetime one has
\begin{equation}
\Phi = {mr^3\over r^4+a^2 z^2} = {m\over r(1+a^2 z^2/r^4)},
\end{equation}
and
\begin{equation}
\ell_a = \left(1 ,  {rx  + ay  \over a^2 + r^2} 
+{ry-ax \over a^2 + r^2} , {z\over r} \right),
\end{equation}
subject to $r(x,y,z)$, which is now a dependent function not a coordinate, being implicitly determined by:
\begin{equation}
\label{E:r-KS}
x^2+y^2+z^2 = r^2 + a^2\left[1-{z^2\over r^2}\right].
\end{equation}
(But this analysis, with suitable substitutions for $\Phi$ and $\ell_i$, will apply to any metric that can be written in Kerr--Schild form). 

\subsection{Cartesian Doran form}

The Doran form of the Kerr metric can be written as~\cite{Kerr-book,Kerr-intro,Doran,River}
\begin{equation}
\fl\qquad
g_{ab} = \eta_{ab} + F^2 V_a V_b + F(V_a S_b + S_a V_b)
= \eta_{cd} (\delta^c{}_a+ FS^c V_a) (\delta^d{}_b + FS^d V_b).
\end{equation}
Here $V$ and $S$ are, (in the background metric $\eta_{ab}$), 4-orthogonal timelike and spacelike unit vectors. In particular $S^0=0$, so that $|S^i|=1$. Furthermore $S^0=0$ also implies that the spatial parts of $S$ and $V$ are 3-orthogonal (in the Cartesian 3-metric). Note that due to the 4-orthogonality of $V$ and $S$ we have $\det(\delta^a{}_{b} + FS^a V_b) = 1$ and that this implies $\det(g_{ab})=-1$.

It is easy to find the inverse metric. First note
\begin{equation}
(\delta^a{}_d - F  S^a V_d ) (\delta^d{}_b + FS^d V_b) = \delta^a{}_b.
\end{equation}
This then implies:
\begin{equation}
\fl \quad
[g^{-1}]^{ab} = \eta^{cd} (\delta^a{}_c - F  S^a V_c )(\delta^b{}_d - F  S^b V_d )
= \eta^{ab} - F^2 S^a S^b - F( S^a V^b + V^a S^b).
\end{equation}
Since $S^0 = 0$ we have  $[g^{-1}]^{00} = \eta^{00} = -1$, the lapse is automatically unity (as it should be for the Doran form of the Kerr metric). 

The 3-projection of the metric and the 3-projection of the inverse metric are now trivially determined (viewed as 3-matrices they are \emph{not} inverses of each other) to be:
\begin{equation}
g_{ij} = \delta_{ij} + F^2 V_i V_j + F[V_i S_j+S_iV_j];
\end{equation}
and
\begin{equation}
[g^{-1}]^{ij} = \delta^{ij} - F^2 S^i S^j - F[V^i S^j+S^iV^j].
\end{equation}
Likewise the shift vector is 
\begin{equation}
\beta^i = [g^{-1}]^{0i} = - F V^0 S^i = - F \sqrt{1+V^2} S^i,
\end{equation}
where we have set $V = |V^i|$, so $V^i = |V| \hat V^i$. 

A suitable orthogonal transformation now yields
\begin{equation}
g_{ij} \sim 
\left[\begin{array}{ccc}1+F^2 V^2 & FV & 0\\ FV & 1 & 0\\0&0&1\end{array}\right];
\end{equation}
and
\begin{equation}
[g^{-1}]^{ij} \sim 
\left[\begin{array}{ccc}1 & -FV & 0\\ -FV & 1-F^2 & 0\\0&0&1\end{array}\right];
\end{equation}
whence
\begin{equation}
\det(g_{ij})=1; \qquad \det([g^{-1}]^{ij})=1-F^2(1+V^2).
\end{equation}

Now determine the susceptibility tensors by 
\begin{equation}
\fl \qquad
\mu^{ij} =  { [g^{-1}]^{ij} \over \det([g^{-1}]^{\circ\circ})};
\qquad
[\epsilon^{-1}]_{ij}  ={g_{ij}\over \det(g_{\circ\circ})};
\qquad
\zeta^{ij} = -{1\over2}\; \left(  \epsilon^{i}{}_{kl}[g^{-1}]^{0l}  [g^{-1}]^{kj}  \right).
\end{equation}

Then
\begin{equation}
[\epsilon^{-1}]_{ij}  =  \delta_{ij} + F^2 V_i V_j + F(V_i S_j+S_iV_j).
\end{equation}
To invert this, perform a suitable orthogonal transformation, then
\begin{equation}
[\epsilon^{-1}]_{ij}  \sim 
\left[\begin{array}{ccc}1+F^2 V^2 & FV & 0\\ FV & 1 & 0\\0&0&1\end{array}\right].
\end{equation}
Invert
\begin{equation}
\epsilon^{ij}  \sim 
\left[\begin{array}{ccc}1 & -FV & 0\\ -FV & 1+F^2V^2 & 0\\0&0&1\end{array}\right].
\end{equation}
Observe $\det(\epsilon_{ij})=1$ and that the eigenvalues of $\epsilon_{ij}$ are always real and positive, though one is greater than unity, one equals unity, and one is less than unity.
Unwrapping the orthogonal transformation
\begin{equation}
\epsilon^{ij}  = \delta^{ij} - F(V^iS^j+S^iV^j) + F^2 V^2 S^iS^j.
\end{equation}
For $\mu$ we simply read off the result
\begin{equation}
\mu^{ij} = { \delta^{ij} - F^2 S^i S^j - F(V^i S^j+S^iV^j)\over 1-F^2(1+V^2)}.
\end{equation}
There is nothing particularly clean one can say about the eigenvalues of $\mu_{ij}$ except that they are either all three positive, or two positive and one negative depending on the sign of the quantity $1-F^2(1+V^2)$. One can establish this by using an orthogonal transformation to write
\begin{equation}
\mu^{ij} \sim { 1\over 1-F^2(1+V^2)} 
\left[\begin{array}{ccc}1 & -FV & 0\\ -FV & 1-F^2 & 0\\0&0&1\end{array}\right].
\end{equation}
Note that $g_{00} = \eta_{00} + F^2 V_0 V_0 = -1 + F^2(1+V^2) = -\{1-F^2(1+V^2)\} $, so the quantity $1-F^2(1+V^2)$ goes through zero and flips sign at the ergo-surface. Indeed two of the eigenvalues of $\mu_{ij}$ diverge at the ergo-surface. (This is similar to what we saw for the Kerr geometry in Kerr--Schild form.)


For $\zeta$ there is a brief computation
\begin{eqnarray}
\zeta^{ij} &=& -{1\over2}\; \left(  \epsilon^{i}{}_{kl}[g^{-1}]^{0l}  [g^{-1}]^{kj}  \right)
\\
&=&
+ {1\over2}\; \left(  \epsilon^{i}{}_{kl} F V^0 S^l  \left[\delta^{kj} + F^2 S^k S^j - F(V^k S^j+S^k V^j)\right]  \right).
\end{eqnarray}
There are a number of cancellations  ($ \epsilon^{i}{}_{kl} S^k S^l = 0$): 
\begin{equation}
\zeta^{ij} 
 =
{FV^0\over2}\; \left(  \epsilon^{i}{}_{kl} S^l  \left[\delta^{kj}  - FV^k S^j\right]  \right)
\end{equation}
Finally
\begin{equation}
\zeta^{ij} 
= {F\sqrt{1+V^2}\over2}\; \left(  \epsilon^{ij}{}_{l} S^l  - FV [\epsilon^{i}{}_{kl} \hat V^k S^l] S^j \right).
\end{equation}

In summary, for any metric in the Doran form elucidated above, the equivalent susceptibility tensors are given by:
\begin{equation}
\epsilon^{ij}  = \delta^{ij} - F(V^iS^j+S^iV^j) + F^2 V^2 S^iS^j.
\end{equation}
\begin{equation}
\mu^{ij} = { \delta^{ij} + F^2 S^i S^j - F[V^i S^j+S^iV^j]\over 1+F^2(1-V^2)}
\end{equation}
\begin{equation}
\zeta^{ij} 
= {F\sqrt{1+V^2}\over2}\; \left(  \epsilon^{ij}{}_{l} S^l  - FV [\epsilon^{i}{}_{kl} \hat V^k S^l] S^j \right).
\end{equation}

If desired, for the Kerr spacetime we can make this fully explicit by inserting the specific form of the metric.
First
\begin{equation}
F = \sqrt{2mr\over r^2+ a^2},
\end{equation}
where $r(x,y,z)$ is the same quantity as appears in the Kerr--Schild Cartesian coordinates. Furthermore
\begin{equation}
V_a = \sqrt{r^2(r^2+a^2)\over r^4+a^2 z^2} \left( 1, {ay\over r^2+a^2}, {-ax\over r^2+a^2}, 0\right),
\end{equation}
and
\begin{equation}
S_a = \sqrt{r^2(r^2+a^2)\over r^4+ a^2 z^2} 
\left( 0, {r x\over r^2+a^2}, {r y\over r^2+a^2}, {z\over r}\right).
\end{equation}
To connect this back to the Kerr--Schild version of Kerr, note $V_a + S_a \propto \ell_a$,
indeed
\begin{equation}
F \; (V_a + S_a ) = \sqrt{\Phi} \; \ell_a.
\end{equation}
(But the current analysis, with suitable substitutions for $F$, $V_i$, and $S_i$, will apply to any metric that can be written in Doran form).

\section{Discussion and conclusions}\label{sec:d}
We have seen above how to choose specific (physically more or less reasonable) laboratory profiles for the permittivity, permeability, and magneto-electric tensors that are suitable for mimicking \emph{at the wave optics level} various coordinate versions of both the Schwarzschild and the Kerr spacetimes. (These being some of the most physically important of the exact solutions in general relativity.) Indeed we have seen how to apply the formalism generally in static spherically symmetric and stationary axisymmetric spacetimes. In principle this process could easily be applied to Reissner--Nordstr\"om and Kerr--Newman (or other more complicated) spacetimes. 

An analysis along these lines is absolutely necessary if one wants to use electromagnetic analogue spacetimes to mimic any specific general relativity spacetime --- once one is presented with the metric of interest in some specific quasi-Cartesian coordinate system, the analysis of this article provides the necessary framework for calculating the appropriate laboratory setup. (If one refuses to use Cartesian coordinates in the laboratory, then one can still make considerable progress, but technical aspects of the calculation are significantly messier. There are situations where this might nevertheless be physically appropriate, 
and work along these lines is ongoing.)

\section*{Acknowledgments}
MV acknowledges financial support from the Marsden Fund administered by the Royal Society of New Zealand.
SS was also supported by a Victoria University of Wellington PhD Scholarship.

\section*{References}
  
\end{document}